\documentclass[letterpaper,10pt,conference]{ieeeconf}
\usepackage{amsmath}
\usepackage{amssymb}
\usepackage{cite}
\usepackage{psfrag}
\usepackage{graphicx}
\IEEEoverridecommandlockouts
\newcommand{\tx}[1]{\text{#1}}

\newcommand{\n}[1]{\vert #1 \vert}

\newtheorem{thm}{Theorem}

\newtheorem{ex}{Example}
\newcommand{\QEDex}{\vspace{-4mm}\begin{flushright}$\diamond$\end{flushright}}
\begin{document}
\title{Improving the Sphere-Packing Bound for Binary Codes over Memoryless Symmetric Channels}
\author{\authorblockN{Kaveh Mahdaviani \authorrefmark{1}, Shervin Shahidi \authorrefmark{2}, Shima Haddadi \authorrefmark{3}, Masoud Ardakani \authorrefmark{1}, and Chintha Tellambura \authorrefmark{1}}\\[5pt]
\authorblockA{\authorrefmark{1} Department of Electrical and Computer Engineering, University of Alberta, Edmonton, AB, Canada, T6G 2V4 \\
Email: \{mahdaviani,ardakani,tellambura\}@ece.ualberta.ca} \authorblockA{\authorrefmark{2} Department of Electrical and Computer Engineering, Queen's University, Kingston, ON, Canada, K7L 3N6 \\
Email: shervin.shahidi@queensu.ca} \authorblockA{\authorrefmark{3}Department of Electrical and Computer Engineering, Isfahan University of Technology, Isfahan, Iran\\
Email: sh.haddadi@ec.iut.ac.ir}}
\maketitle
\begin{abstract}
A lower bound on the minimum required code length of binary codes is obtained. The bound is obtained based on observing a close relation between the Ulam's liar game and channel coding. In fact, Spencer's optimal solution to the game is used to derive this new bound which improves the famous Sphere-Packing Bound.
\end{abstract}
\section{Introduction}
In 1950 Hamming \cite{Hamming1950} introduced the Sphere-Packing Bound (SPB), which gives an upper bound on the number of codewords (i.e., code size) of a block error correcting code of length $n$ and minimum distance $d$. In particular, for a binary block code, we have
\begin{align} \label{eq:Sbin}
S_{\text{bin}}(n) \le \frac{2^n}{\sum_{i=0}^t \binom{n}{i}},
\end{align}
where $S_{\tx{bin}}(n)$ is the size of the code, and
\begin{align} \label{eq:t}
t = \left\lfloor\frac{d-1}{2}\right\rfloor
\end{align}
denotes the error correction capability of the block code. Of course \eqref{eq:t} holds only for the ML decoder \cite{Lin_book_2nd_ed}.For perfect codes \cite{Lin_book_2nd_ed},the inequality \eqref{eq:Sbin} changes to equality. It has been shown that the only known perfect binary block codes are: Hamming code \cite{Hamming1950} for $t = 1, m = 2^i - 1$ for $i\ge3$, and the (23,12) Golay code \cite{Golay49} with  $t = 7$. Different constructions have also been introduced for nonlinear perfect binary codes in the case of $t = 1, m = 2^i - 1, i \ge 3$ \cite{Vasil'ev1962,Etzion1994}. Perfect codes have attracted much interest because of their optimal minimum distance.

\begin{figure}[t!]
\centering
\resizebox{3.4in}{!}{\includegraphics[scale=0.5]{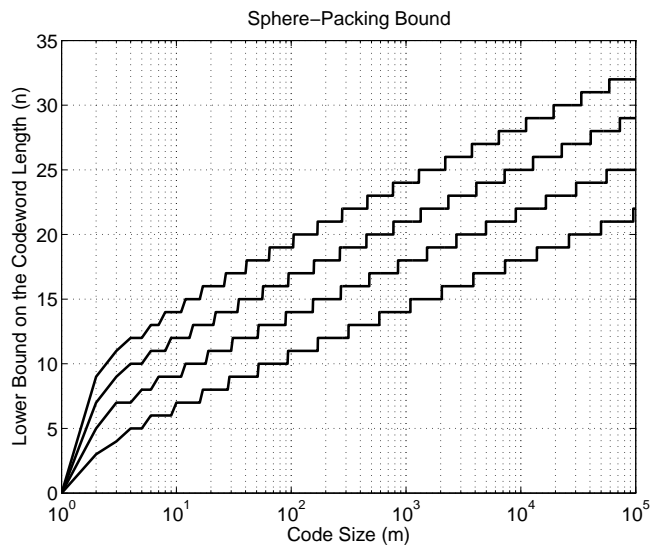}}\\
\caption{Sphere-Packing Bound for binary codes of size $1,\ldots,10^5$ and minimum distance of $3,5,7,9$ (from bottom to top) corresponding to error correction capabilities of $1,2,3,4$  bits.}
\label{fig:SPB}
\end{figure}

Using the SPB, one can easily obtain a curve, which for every pair of integers $m$ and $d$, assigns a lower bound on the required length $n$ of the codewords of a block code of size $m$ and minimum distance $d$. Fig.~\ref{fig:SPB} shows such curves for $m = 1,\dots,10^5$,and $d = 3,5,7,9$ (i.e., $t = 1,2,3,4$).

Using the SPB, for a cannel which does not introduce more than $t$ errors into a codeword, we can find a lower bound on the code length for error-free communication. Unlike this approach, in 1959 Shannon studied a bound on the error probability of a Gaussian channel, where more than $t$ errors could be introduces into a codeword \cite{Shannon1959}. This approach is referred to as improved Sphere-Packing Bound (ISPB) in the literature. Consequently, in 1967 Shannon \emph{et al.} provided another ISPB for discrete memoryless channels \cite{Shannon1967_1,Shannon1967_2}. Valembois and Fossorier improved the latter ISPB in 2004 \cite{Valembois04}. They also extended the result to the binary-input AWGN channel. Recently, in 2008 Wiechman and Sason \cite{Wiechman08} improved the bounding techniques in \cite{Shannon1967_1,Shannon1967_2} and \cite{Valembois04} and derived a new ISPB for all symmetric memoryless channels.

In this article, we would be more faithful to the Hamming's original line and introduce a new improvement in the SPB for binary codes. In other words, instead of being interested about the error probability, we focus on guaranteeing $t$--bit error correction capability.

Since our work is based on Ulam's liar game \cite{Ulam_book_3rd_ed} and its solution, section~II briefly reviews this game and Spencer's optimal solution \cite{Spencer1992}. In section~III, the close relation between Ulam's game and binary channel coding is discussed. Thereafter, in section~IV, using this relation and Spenser's optimal solution, a new upper bound for the maximum size of binary codes is obtained. Comparing this new bound with the famous SPB, it is observed that for some special cases, the new bound is tighter.

\section{Ulam's Game and Spencer State}
\subsection{Ulam's liar game}
Ulam's liar game, which will be referred to as ``U-Game'' in the rest of this work, is a two players game with three parameters $(m,t,n)$. The game starts with Player 1 selecting a symbol among a set $S$ of $m$ different symbols. In order to win, Player 2 must guess the selected symbol with at most $n$ Yes/No questions of the form ``Is the selected symbol among the set $A$?'' where $A$ is a subset of $S$. We will refer to such a question as ``U-Question($A$)''. Throughout the game, Player 1 can give at most $t$ wrong answers. If the Player 2 fails to correctly guess the selected symbol, Player 1 is declared winner.

Hence, Player 2 has to design a series of $n$ U-Questions to deduce the selected symbol. It is important to determine the minimum number of required questions through which one can guarantee that Player 2 wins. If the minimum required number of U-Questions is less than or equal to $n$, Player 2 has a strategy to win the game.

Other variations of the game are also considered in the literature, e.g., \cite{Berlekamp_Thes}, and various solutions have been presented to different versions of the game \cite{Spencer1992,Pelc1987,Deppe2000}.
\subsection{Spencer State Space and Spencer Weight}
Spencer has analyzed the U-Game \cite{Ulam_book_3rd_ed}, where he proposes a state model for the game. Whenever the questioner receives a new answer, this state is updated in a way that it contains all the information which has been received about every symbol up to now. The Spencer's model for an $(m,t,n)$--game, consists of $t + 1$ bins in a row and m chips, $c_1, c_2, \dots , c_m$ corresponding to m symbols. As a result of this one-to-one correspondence, from now on, we use terms ``chip'' and ``symbol'' interchangeably.

In the initial state of the game, all the chips are in the left most bin and the chips are moved to the righter bins according to the received answers. After receiving the $j^{\text{th}}$ answer, the state is denoted by a vector $v_j = (V_0,\dots,V_t)$, where $V_i$ is the subset of the chips in the $i^{\text{th}}$ bin. Notice that the most left bin is indexed zero and the bin index increases to the right. Then the initial state of the game is $v_0 = (S, \varnothing, \dots, \varnothing)$, where $S$ is the set of all chips (symbols).

Now suppose we are at state $v_j$ and the questioner asks the U-Question($A$), where $A$ is a subset of $\{1, \dots, m\}$. Notice that chips corresponding to the elements of $A$ can be in different bins. If the answer to this question is a ``No'' we update the state by moving all the chips corresponding to the elements of $A$ one bin to the right. A chip moving to the right of the right-most bin is considered ``\textit{lost}''. If the answer is a ``Yes'' we can view it as a ``No'' to U-Question($A^c$) and use the mentioned update rule.

With the initial state $v_0$ and this updating process, it is evident that a chip $c_i$ will be lost if and only if the questioner receives more than $t$ answers stating that $c_i$ is not corresponding to the selected symbol. Thus, $c_i$ cannot be the selected symbol by Player 1. Obviously, Player 2 wins the game if within n questions he observes a state where all chips except one are lost.

To simplify the analysis of the game, we define notations to present the above discussion. To this end, we denote the set of symbols belonging to $A$ in the $i^{\text{th}}$ bin by $U_i$. We now view $A$, which is the matter of question at step $j + 1$, by vector $u_{j+1} = (U_0, \dots, U_t)$.  Then, according to the above mentioned update rules, we can represent the updated state in the case of receiving a ``No'' as
\begin{align} \label{eq:No}
v_{j+1} &= \tx{No}\{v_j,u_{j+1}\}  \\
&\triangleq ((V_0 \setminus U_0), (V_1 \setminus U_1) \cup U_0, \dots, (V_t \setminus U_t) \cup U_{t-1}), \nonumber
\end{align}
and in the case of ``Yes'' as
\begin{align} \label{eq:Yes}
v_{j+1} &= \tx{Yes}\{v_j,u_{j+1}\}  \\
&\triangleq ((U_0, U_1 \cup (V_0 \setminus U_0), \dots, U_t \cup (V_{t-1} \setminus U_{t-1})). \nonumber
\end{align}

Spencer has also introduced a weight for every state of the game. The weight of a state $v_j = (V_0, \dots, V_t)$, is defined as
\begin{align}\label{eq:Weight}
W(v_j) \triangleq \sum_{i=0}^t \left[\n{V_i} \sum_{\ell=0}^{t-i} \binom{n-j}{\ell}\right].
\end{align}

We would refer to this weight function as ``Spencer weight''. Spencer showed that if in any step $i$ through the game, the state weight is greater than $2^{(n-i)}$, then there is surely a strategy for Player 1 to win \cite{Spencer1992}.

\section{The Relation Between Channel Codes and the Ulam's Game}
The main problem in binary error correction coding is very similar to a U-Game. To transmit $\log_2 m$ information bits, the transmitter selects a symbol from a set of cardinality $m$, and then sends a series of n bits (0 or 1) through the channel in order to inform the receiver what symbol has been selected The channel then flips some of the bits and the receiver should use the received bits to deduce the selected symbol.

The aim of a $t$-bit error correcting code is to guarantee correct decoding if the channel has flipped at most $t$ bits. The main problem here is again to design a code with minimum possible length to guarantee the $t$-bit error correction capability. In a well designed decoder, the parameter $t$ is related with the minimum distance of the codewords as in \eqref{eq:t}.

According to the $i^{\text{th}}$ bit of the codewords, the codebook can be partitioned into two sets. The set $A$ of all codewords whose $i^{\text{th}}$ bit is `1', and $A^c$ of all codewords whose $i^{\text{th}}$ bit is `0'. Thus, a block code of length $n$ can be viewed as a series of $n$ U-Questions\footnote{Here, a block code of length $n$ is considered. The discussions, however, are valid for the case of variable-length codes.}. The channel can give incorrect answers to some of these $n$ questions.

There exists, however, a few differences between the two problems. The first and the most important one is that in the coding case, the questions are preset, i.e., the codebook is designed before transmission. In the case of the U-Game, however, Player 2 can use the answers received up to now to design the future questions.
The second difference is that in Ulam's game, Player 1 can choose a lying strategy to make deduction of the selected symbol harder, while channel errors occur randomly. In other words, errors are not planned by the channel.

Since the goal of the coding problem is to guarantee an error correction capability of $t$ bits, one should consider the worst case errors. Thus, without loss of generality, we can assume that the channel errors are planned to make the decoding harder. Therefore, channel coding can be viewed as a U-Game where Player 1 (channel) is still playing based on its best strategy, while Player 2 (code designer) must design all his questions at the beginning of the game. Thus, the minimum number of questions, required in an $(m,t,n)$--game is a lower bound on the minimum required length of a code of size $m$ with error correction capability $t$. For channels with real-time feedback, code designer can use the best strategy available to Player 2 in U-Game making both problems identical from this point of view.

Another minor difference between the two problems is that the channel does not care about the maximum allowed number of incorrect answers. That is, it may introduce more than $t$ errors. In such cases, the decoder fails. This failure, however, does not have any effect on the code design because our code is only concerned about guaranteeing successful decoding when the number of errors is no more than $t$. Thus, in the sequel, we limit our discussions to the cases that no more than $t$ errors are occurred.

The following theorem relates failure of channel decoding to the Spencer weight of the last state of the equivalent U-Game.
\vspace{3mm}
\begin{thm} \label{thm:endgame}
In a communication system, equipped by a $t$--bit error correcting code of length $n$ and size $m$, if at the end of the equivalent $(m,t,n)$--game the Spencer weight is greater than one, there is no guarantee of successful decoding.
\end{thm}
\vspace{3mm}
\begin{proof}
To proof this theorem, we get advantage of the optimality of Spencer's solution, in the means of minimum required questions. From the definition of the Spencer weight in \eqref{eq:Weight} we have the Spencer weight of a Spencer state $v_n = (V_0, \dots, V_t)$ at the end of the equivalent $(m,t,n)$--game as
\begin{align} \label{eq:LastWeight}
W(v_n) = \sum_{i=0}^t \left[\n{V_i} \sum_{\ell=0}^{t-i} \binom{n-n}{\ell}\right]>1,
\end{align}
and since
\begin{align} \label{eq:binom_0_n}
\sum_{\ell=0}^{t-i} \binom{n-n}{\ell}=1,\; \forall~ t-i,n\in\mathbb{N},
\end{align}
then
\begin{align} \label{eq:LastWeight_concloud}
W(v_n) = \sum_{i=0}^t \n{V_i}>1.
\end{align}

Where in \eqref{eq:LastWeight_concloud}, the left side of the inequality is the number of the chips remained in the state at the end of the equivalent game supposing we have used Spencer's method to solve it. This situation means that the information received by the transmitted bits, is not enough to deduce which message have been selected in the transmitter. In such cases, although the receiver may select one of the possible blocks, but there will be no guarantee on the correctness of this decoding.
\end{proof}

\section{The New Bound}
In this section, based on the relation between U-Game and the channel coding problem, we use Spencer's optimal solution in order to obtain a lower bound on the codeword length. In other words, we find a bound on the required number of bits to describe a selected symbol from a set of $m$ predefined symbols, when at most $t$ bits could be received incorrectly.  We then observe that this bound is slightly tighter than the well known SPB.

Let us first go through a simple example, where the lower bound obtained by SPB could be improved using Spencer's solution.
\vspace{1mm}
\begin{ex}
For a set of three symbols to be transmitted through a channel using binary error correcting codes and guaranteeing the correction of every error of hamming weight one, the Sphere-Packing Bound gives us a lower bound of four on the minimum required length of codewords as

\begin{figure}[t!]
\centering
\psfragscanon
\psfrag{c1}{{$c_1$}}
\psfrag{c2}{{$c_2$}}
\psfrag{c3}{{$c_3$}}
\footnotesize
\psfrag{A1}{$A_1 = \{c_1,c_2\}$}
\psfrag{x1}{$x_1=1$}
\psfrag{A2}{$A_2 = \{c_1,c_3\}$}
\psfrag{x2}{$x_2=1$}
\psfrag{A3}{$A_3 = \{c_1\}$}
\psfrag{x3}{$x_3=0$}
\psfrag{A4}{$A_4 = \{c_1,c_2\}$}
\psfrag{x4}{$x_4=1$}
\psfrag{A5}{$A_5 = \{c_1\}$}
\psfrag{x5}{$x_5=1$}
\resizebox{3.4in}{!}{\includegraphics[scale=0.56]{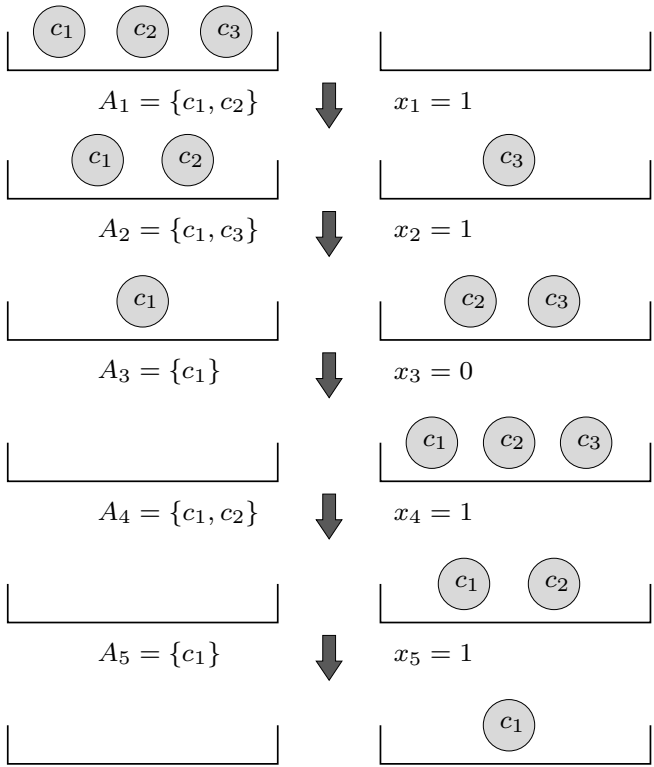}}\\
\caption{The selected symbol $c_1$ is deduced through five U--Questions while a wrong answer (at step three) has made the state of the game inconclusive after the fourth answer. In this figure $A_i$, $i = 1,\dots,5$ represents the subset under question in U--Questions $1,\dots,5$, and $x_i$, $i = 1,\dots,5$ represents the received answer (or bit) where `1' means ``Yes'' and `0' means ``No''}
\label{fig:example}
\end{figure}

\begin{align} \label{eq:Example}
4 = \min_{x\in\mathbb{N}}\left\{x\bigg\vert 3\le\frac{2^x}{\sum_{i=0}^1 \binom{x}{i}}\right\}.
\end{align}

But now, let $c = [b_0 b_1 b_2 b_3]$ be a codeword. To be able to correct every error of Hamming weight one, we need then to have a Hamming distance of at least three between every pair of codewords.  It is, however, easy to check that there exists no pair of vectors with Hamming distance three or more among all vectors of distance at least three from $c$, i.e., $[\bar{b_0} \bar{b_1} \bar{b_2} b_3]$, $[\bar{b_0} \bar{b_1} b_ 2 \bar{b_3}]$, $[\bar{b_0} b_1 \bar{b_2} \bar{b_3}]$, and $[\bar{b_0} \bar{b_1} \bar{b_2} \bar{b_3}]$. Thus, the lower bound provided by SPB cannot be achieved by any error correcting code of size three and length four. On the other hand, if we think of the equivalent $(3,1,n)$--game, as shown in Fig.~\ref{fig:example}, for $n = 4$ the game is not conclusive. In other words, after the fourth question, we still have two chips left in the game. One can easily check that these U-Questions are the bests, and the received answers are the worst. Thus, a bound on minimum codeword length $n$ for $m = 3$, $t = 1$ can be obtained from this U-Game to be $n \ge 5$. Interestingly, a code with $n = 5$ can  in fact be constructed for example with codewords $c_1=[00000]$, $c_2=[11100]$, $c_3=[11011]$.
\QEDex
\end{ex}

Through the rest of this section, we introduce a mathematical framework to obtain a new lower bound on the codeword length, using its equivalent U-Game. Before going through more details and formulating the new improved bound, we introduce some definitions.

For an error correcting code with length $n$ and size $m$ and error correcting capability $t$, let
\begin{align} \label{eq:A_n-s}
A_{n-s} \triangleq \gcd\left\{{\binom{n-s}{t}, \dots, \binom{n-s}{t-s+1}}\right\},
\end{align}
and
\begin{align} \label{eq:K_0}
K_0 \triangleq m \times \sum_{\ell=0}^{t}{\binom{n}{\ell}}.
\end{align}
Then we calculate $K_i$ recursively from $K_{i-1}$ using the following rule:  $K_i$ should be the least integer satisfying
\begin{align} \label{eq:K_i_half}
K_i \ge \frac{K_{i-1}}{2},
\end{align}
and
\begin{align} \label{eq:K_i_mod}
K_i \equiv m \times \sum_{\ell=0}^{t}{\binom{n-i}{\ell}}\; (\tx{mod } A_{n-i}).
\end{align}

Now we introduce a new bound through the next theorem.
\vspace{3mm}
\begin{thm} \label{thm:main}
A code of length $n$, size $m$, and error correction capability $t$ exists if for all $1\le i\le n$,
\begin{align} \label{eq:Theo_main}
K_i \le 2^{n-i}.
\end{align}
\end{thm}
\vspace{3mm}
\begin{proof}
To prove this theorem, we show that $K_i$ is less than or equal to the Spencer weight of the equivalent U-Game after the $i^{th}$ U-Question is answered. Thus, if $K_n > 1$, the Spencer weight after the $n^{th}$ answer is also greater than 1. Therefore, according to Theorem~\ref{thm:endgame}, the game is inconclusive.

To show that $W(v_i) \le K_i$, we notice that initially the Spencer weight of the equivalent U-Game is exactly equal to the $K_0$ by the definition. Then after each update the new Spencer weight should have three conditions. First, it should be an integer, since as defined in \eqref{eq:Weight} the Spencer weight is a summation in which every term is the product of the number of chips in a bin and a combination term, which are both integers. The second condition as we will show is that after each update, the maximum guaranteed reduction in the Spencer weight is half of the weight. In other words, if we consider the worst case by the means of the least possible reduction in the Spencer weight, then we have
\begin{align} \label{eq:W_i_half}
W(v_i) \ge \frac{W(v_{i-1})}{2}.
\end{align}

In order to show this condition holds, suppose we are in an arbitrary Spencer state, $v_{i} = (V_0, \dots, V_t)$ in the equivalent $(m,t,n)$ U-Game, and we are going to ask the U-Question($A$) where $A$ could be described by the vector $u_{i+1} = (U_0, \dots, U_t)$. Regardless of the question, the updated Spencer state is either $\text{Yes}\{v_{i},u_{i+1}\}$ or $\text{No}\{v_{i},u_{i+1}\}$. Then Using \eqref{eq:Yes}, \eqref{eq:No}, and \eqref{eq:Weight}, the sum of the Spencer weight of the two possible results is
\begin{align} \label{eq:Y_N_Sum}
W(\text{Yes}\{v_{i},u_{i+1}\}) + W(\text{No}\{v_{i},u_{i+1}\}) = W(v_{i}).
\end{align}

Hence we have,
\begin{align} \label{eq:Max_Min}
\max_{u_{i+1}}\left\{\min\{W(\text{Yes}\{v_{i},u_{i+1}\}), W(\text{No}\{v_{i},u_{i+1}\})\}\right\} \\  = \frac{W(v_{i})}{2}. \nonumber
\end{align}
Here maximization is taken over all possible questions.

The third condition is that regardless of what the answer of a question is, the Spencer weight of the new states must satisfy the following condition:
\begin{align} \label{eq:W_i_mod_A_n-i}
\forall i \le n,\; m \times \sum_{\ell=0}^{t}\binom{n-i}{\ell} \equiv W(v_i)\; (\tx{mod }{A_{n-i}})
\end{align}
which is proved in \cite{Spencer1992}.

As a result, $K_i \le W(v_i)$ and \eqref{eq:Theo_main} can be used to obtain a lower bound on $n$.
\end{proof}
\vspace{3mm}

\begin{figure}[t!]
\centering
\resizebox{3.4in}{!}{\includegraphics[scale=0.5]{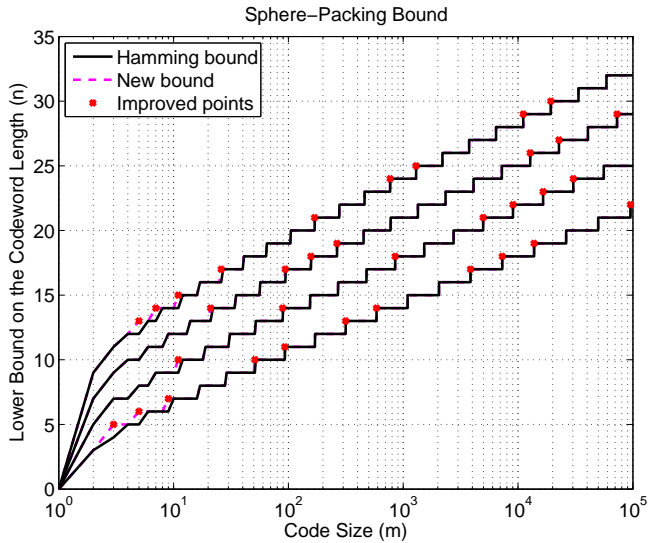}}\\
\caption{The Sphere-Packing Bound versus the new bound for binary codes of size $1,\ldots,10^5$ and error correction capability of $1,2,3,4$ bits. The solid black lines are the famous SPB introduced by Hamming, which are the same as in Fig.~\ref{fig:SPB}. The dashed red lines are the new bound with respect to Theorem~\ref{thm:main}.}
\label{fig:newbound}
\end{figure}

The following theorem states that the new bound is at least as good as the famous SPB.
\vspace{3mm}
\begin{thm} \label{thm:better}
For any $m$ and $t$, the lower bound on $n$ obtained based on Theorem~\ref{thm:main} is at least as tight as the Hamming bound.
\end{thm}
\vspace{3mm}
\begin{proof}
Assume that for some $m$ and $t$, a bound $n$ looser than the Hamming bound is obtained through Theorem~\ref{thm:main}. Then, using the pigeon-hole principle,  at least two of the $m$ spheres with radius $t$, centered at $m$ codewords, will intersect.

Recall that any point in this space can be viewed as a sequence of answers in the equivalent U-Game. Now, if Player 1 picks one of the centers of these two intersecting spheres and answers the questions according to one intersecting point, Player 2 will be left with more that one choice at the end of the game. This is because both centers of the intersecting spheres are less that $t$ apart from the given sequence of answers. Thus, Player 1 with at most $t$ wrong answers can win the game.

Since Theorem~\ref{thm:main} guarantees existence of a winning strategy for Player 2 \cite{Spencer1992}, the assumption that the new bound can be looser than the Hamming bound is contradicted.
\end{proof}
\vspace{3mm}

Fig.~\ref{fig:newbound} shows a comparison between the famous SPB and the bound achieved by Theorem~\ref{thm:main}. It contains four pairs of curves for $t = 1, 2, 3, 4$ from bottom to top, respectively. As we can see, the two bounds are usually the same. However, in some particular cases, which are shown by solid circles, the new bound describes a tighter lower bound on the minimum number of required bits. As predicted by Theorem~\ref{thm:better}, the new bound is never looser than SPB. Indeed, the fact that the new bound is at least as tight as SPB can be used to reduce the computational complexity of finding the new bound. To this end, one can use SPB as a starting point to search for the smallest $n$ satisfying Theorem~\ref{thm:main}.

\section{Conclusion} \label{sec:conlusion}
In this paper we first discussed the relation between the error correcting codes and the Ulam's game. Then we discussed that any binary error correcting code has an equivalent U-Game. Finally, using Spencer's solution to U-Game, we derived a new lower bound on the minimum length of the codewords of an error correcting code of size $m$ and error correction capability $t$. The new bound was proved to be at least as tight as SPB and was shown to be better than the famous Sphere-Packing Bound is some cases.
\bibliographystyle{IEEEtran}
\bibliography{IEEEabrv,kaveh_bib}
\end{document}